\documentclass[12pt]{article}
\usepackage{putex}
\usepackage{graphicx}
\def \ie {{\it i.e.}}
\def \g {g_{\rm str}}
\def \sp {\mbox{ }}

\begin{document}

\preprint{hep-th/0503193 \\ PUPT-2156}

\institution{PU}{Joseph Henry Laboratories, Princeton University, Princeton, NJ 08544}

\title{Instabilities of D-brane Bound States and Their Related Theories}

\authors{Joshua J. Friess and Steven S. Gubser}

\abstract{We investigate the Gregory-Laflamme instability for bound states of branes in type~II string theory and in
M-theory.  We examine systems with two different constituent branes: for
instance, D3-F1 or D4-D0.  For the
cases in which the Gregory-Laflamme instability can occur, we describe the
boundary of thermodynamic
stability.  We also
present an argument for the validity of the Correlated Stability Conjecture, generalizing earlier work by Reall.  We discuss the
implications for OM theory and NCOS theory, finding that in both cases, there exists some critical temperature above
which the system becomes unstable to clumping of the open strings/membranes.}


\date{November, 2005}

\maketitle

\tableofcontents

\section{Introduction}
\label{INTRODUCTION}

The Gregory-Laflamme (GL) instability \cite{glOne,glTwo} is expected to arise on extended black hole horizons when there is a non-uniform state with the same mass and charges but greater entropy.  A precise version of this expectation is the Correlated Stability Conjecture (CSC) \cite{gmOne,gmTwo}, which says that a GL instability should arise when the matrix of second derivatives of the mass with respect to entropy and conserved charges fails to be positive definite.  There is a partial proof of the CSC \cite{reall}, and no fully demonstrated counter-examples have appeared to our knowledge.

In recent work \cite{ssg-recent}---to which we refer the reader for a more complete list of references---it was
shown that the CSC holds in the case of the D2-D0 bound state.  For that system, there is a stability curve in the space of
solutions: for a given number of D2-branes and number density of D0-branes, there is a certain
non-extremal mass (\ie, temperature) above
which the GL instability occurs and below which it does not.  This stability curve comes arbitrarily close to extremal
solutions as one increases the number density of D0-branes.  The infinite density limit describes a non-commutative
field theory, so one
learns from the GL instability that there is a subtlety in the NCFT limit: the unstable mode corresponds to developing non-uniform non-commutativity, and it was suggested that this unstable mode runs away to long wavelength in the NCFT limit so as to avoid conflict with standard decoupling arguments.

In section~\ref{CASES} we generalize the thermodynamic analysis of \cite{ssg-recent} for other D-brane bound states
in type~II string theory and M-theory.  We do not repeat the numerical analysis of \cite{ssg-recent} for these cases;
instead, in section~\ref{ANALYTICAL}, we revisit the analytical arguments of \cite{reall} for the validity of the
CSC.  These arguments (at least, as presented here) are not completely airtight, but they strongly suggest that the
CSC is valid for
the cases we consider.  In section~\ref{OMNCOS} we discuss applications to NCOS theory
\cite{gukov,seiberg,gopakumar} and OM theory \cite{om}.

While this paper was in preparation, we received \cite{Ross:2005vh}, which overlaps with the present work.

\section{Dynamical and Thermodynamic Instabilities}
\label{ANALYTICAL}

In \cite{reall}, an argument was offered in support of the CSC when the entropy is the only
locally dynamic quantity.  Here we will give some indications on how to extend that argument to
the general case, where both charges and angular momenta can be locally manipulated.  We
emphasize that the arguments presented here are more of an outline than a proof, with attention
drawn to the gaps in the reasoning.  The CSC has proven remarkably robust, but should a
counter-example arise, it will be interesting to see where the line of argument presented here
fails.

A central object is the Euclidean action reduced along the direction of the instability.  Let us
call this action $I$; let us parametrize the direction of the instability by the coordinate $y$;
and let $x^\mu$ be the other coordinates, so that $I$ is an integral over the $x^\mu$.  It is
simplest to consider the case where $y$ is the only spatial direction along which the horizon is
extended, so that is the case we will focus on here.  Including more directions of spatial extent
is not a significant obstacle.

The first step is to identify the Hessian matrix of $I$ with the Hessian matrix of
susceptibilities, up to some change of basis and an overall positive factor, so that the signs of
all eigenvalues (as well as their ratios) are preserved.  Let the $n$ quantities $B_i$ be
intensive thermodynamic variables: $B_0$ is the temperature $T=1/\beta$, and the other $B_i$ are
gauge potentials $\mu$ at the horizon and angular velocities $\Omega$ at the horizon.  Let $A_i$
be the thermodynamic conjugate variables that enter into the first law:
 \eqn{FirstLaw}{
  dE = \sum_i B_i dA_i \,.
 }
So $A_0=S$, the entropy, and the other $A_i$ are charges and angular momenta.  The energy $E$ and
the $A_i$ with $i>0$ are conserved quantities, determined by Smarr type integrals at infinity.  It is a fairly generic circumstance that there exists an $n$-parameter family of configurations, labeled by parameters $a_i$, such that
 \eqn{SpecialAction}{
  I = \beta E(a_i) - S(a_i) - \sum_{j>0} \beta B_j A_j(a_i) \,.
 }
The on-shell configuration yielding a uniform brane corresponds, by convention, to $a_i=0$.  The quantities $E(a_i)$, $S(a_i)$, or $A_j(a_i)$ for arbitrary $a_i$ are defined by the same integrals as for on-shell configurations, as is the action $I$.  A further discussion of why we expect there to be an $n$-parameter family of configurations with the properties described is given in the appendix.  But the discussion is incomplete, and this point is one of the aforementioned gaps in the reasoning.

The action must be stationary at $a_i=0$ with respect to variations
of $a_i$.  This leads to
 \eqn{GetThermo}{
  {\partial I \over \partial a_i} &=
    \beta {\partial E \over \partial a_i} -
    {\partial S \over \partial a_i} -
    \sum_{j>0} B_j {\partial A_j \over \partial a_i} = 0  \cr
  {\partial E \over \partial a_i} &=
   \sum_j B_j {\partial A_j \over \partial a_i} \,.
 }
The second line follows from the first, and it is clearly an instance of the first law \FirstLaw.
Imposing this equation for all values of $a_i$ leads to definite expressions for $B_j$ in terms
of the $a_i$.  Differentiating then leads to
 \eqn{DifferentiateAgain}{
  {\partial^2 E \over \partial a_i \partial a_k} &=
   \sum_j \left( {\partial B_j \over \partial a_k}
     {\partial A_j \over \partial a_i} +
    B_j {\partial^2 A_j \over \partial a_i \partial a_k} \right) \,.
 }
On the other hand, differentiating the first line in \GetThermo\ gives
 \eqn{FinalDerivative}{
  {\partial^2 I \over \partial a_i \partial a_k} &=
    \beta {\partial^2 E \over \partial a_i \partial a_k} -
    \beta \sum_j B_j {\partial^2 A_j \over \partial a_i \partial a_k}
   = \beta \sum_j {\partial B_j \over \partial a_k}
    {\partial A_j \over \partial a_i}  \cr
   &= \beta \sum_{j,\ell} {\partial A_\ell \over \partial a_k}
    {\partial B_j \over \partial A_\ell}
    {\partial A_j \over \partial a_i}
   = \beta \sum_{j,\ell} {\partial A_\ell \over \partial a_k}
    {\partial^2 E \over \partial A_\ell \partial A_j}
    {\partial A_j \over \partial a_i}
 }
where in the final step we once again go to on-shell configurations and use $B_j = \partial E /
\partial A_j$.  Comparing the first and last forms in \FinalDerivative, we see that (provided
$\partial A_j / \partial a_i$ is non-singular at $a_i=0$) the Hessian matrix of $I$ with respect to a
specific set of perturbations is equal, up to a positive factor and a change of basis, to the
Hessian matrix of susceptibilities.

The second step is to consider normalizable on-shell perturbations of the extended horizon.  Let
us collectively denote the field perturbations as $\Psi^I$: these fields include the metric
perturbation $h_{MN}$ as well as perturbations in the matter fields.  The ansatz for a static GL
perturbation (still in Euclidean signature) is $\Psi^I(x,y) = \Re(\psi^I(x) e^{iky})$.  Then,
assuming a two-derivative action, the linearized equations of motion take the form
 \eqn{PerturbEOMS}{
  \int d^d x' \, {\delta^2 I \over \delta\psi^I(x) \delta\psi^J(x')}
    \psi^J(x') = -k^2 G_{IJ} \psi^J
 }
for some metric $G_{IJ}$.  Define the $L^2$-like norm $||\psi||^2 = \int d^d x \, \psi^I G_{IJ}
\psi^J$.  Unfortunately, this norm is not positive definite: in pure gravity, a conformal metric
perturbation has $\psi^I G_{IJ} \psi^J < 0$ everywhere.  If one demonstrates the existence of a
perturbation with a positive norm and positive $k^2$, then a GL instability should exist.  The
reasoning is that by taking $k^2$ slightly smaller than the value that gives a static
perturbation, and assigning instead some time-dependence $e^{\pm i \omega t_E}$, one will obtain
a mode which grows exponentially in Minkowskian time.

In \cite{reall}, the on-shell modes of interest were demonstrated through a constraint equation
to have $\psi^I G_{IJ} \psi^J > 0$ everywhere.  The demonstration is somewhat detailed, and
generalizing it rigorously to the cases of interest is another significant gap in the reasoning.
Assuming that the on-shell perturbations $\psi^I$ are indeed restricted to the positive-norm
sector, one sees from \PerturbEOMS\ that directions of quadratically decreasing $I$ are
associated with a GL instability:
 \eqn{GLlink}{
  I - I_0 = {1 \over 2} \int d^d x \, d^d x' \,
   \psi^I(x)
    {\delta^2 I \over \delta\psi^I(x) \delta\psi^J(x')}
   \psi^J(x') + O(\psi^3) =
    -{k^2 \over 2} ||\psi||^2 + O(\psi^3)
 }
where $I_0$ is the action of the original on-shell configuration.

The final step is to argue that there is a direction in which $I$ decreases quadratically with
perturbation fields $\psi^I$ of positive, finite norm precisely when $\partial^2 I/\partial a_i
\partial a_j$ fails to be positive definite.  This is a plausible but non-trivial claim: the
$\psi^I$ perturbations appropriate to a GL instability have not been shown to overlap
sufficiently with the perturbations generated by making the $a_i$ slightly different from the
$A_i$ of the original on-shell configuration.  This is a third gap in the reasoning.

In summary, the core arguments of \cite{reall} generalize to a plausible line of reasoning for
why the CSC is true.  The first and third gaps in the reasoning identified above basically come
down to establishing, at least to second order in the $a_i$, the existence of an
$n$-parameter family of off-shell configurations whose action can nevertheless be written in a
first law form, as in \SpecialAction, and which differ from the original on-shell configuration
by perturbations which have finite positive norm.  The second gap in the reasoning is to show
that the on-shell perturbations for a static mode of finite wavelength have positive norm
(avoiding the conformal factor problem).

The loopholes we have pointed out suggest that perhaps a counter-example to the CSC might be found by setting up a Lagrangian
with special properties.  But the numerics in \cite{ssg-recent} seem to us good preliminary evidence that the CSC does hold
for type II supergravity.  In the remainder of the paper, we will apply it to some interesting cases.

\section{Thermodynamics of D-brane Bound States}
\label{CASES}

The aim of this section is to explore the bound states in type II string theories and eleven-dimensional supergravity
that involve only two types of BPS branes.  We restrict attention to the cases where the worldvolumes of the
lower-dimensional branes are entirely contained in the worldvolume of the higher-dimensional branes.
These cases are:
 \begin{itemize}
  \item D$p$-D($p-2$), explored in section~\ref{DPTWO}.
  \item D$p$-F1, explored in section~\ref{DPFONE}.
  \item D$p$-D($p-4$), explored in section~\ref{DPFOUR}.
  \item M5-M2, which is equivalent for our purposes to D4-F1.
 \end{itemize}
One could also investigate branes intersecting at angles: for instance, one stack of D2-branes extended over the $12$ directions and smeared over the $34$ directions, intersecting another stack extended over $34$ and smeared over $12$.  T-dual relationships, like the relation of the intersecting D2-branes with D4-D0, or the relation of D2-D0 to D3-D1, do not imply identical GL instabilities.  The reason is that the GL instability is an infrared effect, so if one compactifies along the direction of the instability in preparation for taking a T-duality, the instability may be lost.  The interplay of T-duality and the GL instability has recently been understood in some detail for the simple case of smeared D0-branes in \cite{Bostock:2004mg,AharonyEtAlOne,Harmark:2004ws}.

\subsection{The D$p$-D($p-2$) Case}
\label{DPTWO}

The thermodynamic quantities of the D$p$-D($p-2$) bound state with zero angular momentum are given by \cite{harmark}
 \eqn{thermo}{
  M &= {V_p \Omega_{8-p} \over 16 \pi G_N} \sp
    r_0^{7-p} (8 - p + (7 - p) \sinh^2\alpha)  \cr
  T &= {7-p \over 4\pi r_0 \cosh\alpha} \qquad
    S = {V_p \Omega_{8-p} \over 4 G_N} \sp r_0^{8-p} \cosh\alpha
    \cr\noalign{\vskip2\jot}
  \mu_p &= \mu \cos\theta \qquad Q_p = Q \cos\theta \qquad
    \mu_{p-2} = \mu \sin\theta \qquad Q_{p-2} = Q \sin\theta  \cr
  \mu &= \tanh\alpha \qquad
    Q = {(7-p) V_p \Omega_{8-p} \over 16 \pi G_N} \sp r_0^{7-p}
     \sinh\alpha\cosh\alpha \,,
 }
where $\Omega_{8-p}$ is the volume of a unit $S^{8-p}$.  Here it is assumed that the D$(p-2)$ brane is embedded in the
D$p$-brane and smeared along its two transverse directions.

Calculating the boundary of thermodynamic stability is fairly straightforward.  Our system will be locally thermodynamically
unstable if the Hessian of $M$ with respect to any spatially varying quantities has a negative eigenvalue.  In this case, we
take the D$p$-branes to form a static background on which the D$(p-2)$'s can move.  Hence we want to determine the
first and second derivatives of $M$ with respect to $Q_{p-2}$ and $S$, holding $Q_p$ fixed.

In \thermo, $M$ is not expressed as a function of $S$, $Q_p$, and $Q_{p-2}$; instead, these four quantities are expressed in
terms of $r_0$, $\alpha$, and $\theta$.  It is useful to recall a fact from multi-variable calculus: if there is a
smooth, invertible relationship between $n$ variables $Q_i$ and $n$ other variables $q_i$, and $M$ is known as a
smooth function of
the $q_i$, then
 \eqn{NiceFact}{
  \left( {\partial M \over \partial Q_i} \right)_{Q_j} =
   {\partial(Q_1,\ldots,\hat{Q}_i,M,\ldots, Q_n)/
     \partial(q_1,\ldots,q_n) \over
    \partial(Q_1,\ldots,Q_n)/\partial(q_1,\ldots,q_n)} \,,
 }
where the denominator is the Jacobian, $\det(\partial Q_i/\partial q_j)$, and the hat notation in the numerator is meant to
indicate replacing $Q_i$ with $M$.  Furthermore, the thermodynamic dual quantities $T$, $\mu_p$, and $\mu_{p-2}$ should be
precisely the first derivatives of $M$ with respect to $S$, $Q_p$, and $Q_{p-2}$ because of the first law of thermodynamics:
 \eqn{FirstLawTwo}{
  dE = TdS + \mu_p dQ_p + \mu_{p-2}\sp dQ_{p-2} \,.
 }
This can be verified explicitly using \thermo.  Hence, the Hessian may be expressed as $H =
\partial(T,\mu_{p-2})/\partial(S,Q_{p-2})$, and the right hand side may be evaluated by use of \NiceFact.  We find
 \eqn{DetH}{
  \det H = { 16 G_N^2 \sech^4\alpha \over \Omega_{8-p}^2\sp
   r_0^{16-2p}\sp V_p^2 \left((9-p) \cosh^2\alpha -1\right)} \left((5-p)\sinh^2\alpha \cos^2\theta - 1\right) \,.
 }
One can show that this expression reduces to the D2-D0 case studied in \cite{ssg-recent}.  Subsequently, the
condition for thermodynamic {\it stability} is
 \eqn{FSC}{
  \csch\alpha < \sqrt{5-p}\sp \cos\theta \,.
 }
As a check on this result, we note that at extremality ($\alpha \to \infty$) the left hand side vanishes, and the system is
stable as expected.

This stability condition can be recast in terms of the potentials $\mu_p$ and $\mu_{p-2}$ appearing in \thermo\ as
 \eqn{FSCagain}{
  \mu_{p-2}^2 + (6-p) \mu_p^2 = 1 \,.
 }
Using string dualities, we can also immediately recover results for three other cases of interest.  The D3-D1 bound
state is related by S-duality to the D3-F1 bound state, and hence the boundary of stability is the same.
Furthermore, since the D4-D2 bound state is a simple compactification of the M5-M2 state, the above stability
conditions are valid for M5-M2 for $p=4$ (see also \cite{Harmark:2000ff} for an explicit study of the non-extremal M5-M2 bound state and its relation to OM theory).  Furthermore, since we can instead choose to compactify M-theory on a
longitudinal direction of both the M5 and M2 branes, the $p=4$ case must also be the result for
the D4-F1 bound state.

We emphasize again that T-duality acts non-trivially on the CSC rules: indeed, the D3-D1 and D2-D0 cases are related
by T-duality, but their stability conditions are different.

\subsection{The D$p$-F1 Case}
\label{DPFONE}

The extremal supergravity background of the D$p$-F1 bound state was given in \cite{roy}.  The non-extremal generalization \cite{Harmark:2000wv} is
\eqn{DpFone}{
  ds_{\rm str}^2 &= \left({H \over D}\right)^{1/2} \Bigg[
   H^{-1}\left(-f dt^2 + dx_1^2\right) + \left({H \over D}\right)^{-1}
    \left(dx_2^2 + \cdots + dx_p^2 \right)
    \cr &\qquad\quad{} +
    f^{-1} dr^2 + r^2 d\Omega_{8-p}^2 \Bigg]  \cr
  e^\phi &= e^{\phi_0} H^{(3-p)/4} D^{(p-5)/4} \cr
  B_2 &= \sin\theta \coth\alpha \left(1 - H^{-1}\right) dt \wedge dx_1
}
where
 \eqn{HDF}{
    H(r) = 1 + {r_0^{7-p} \sinh^2 \alpha \over r^{7-p}} \qquad
    D(r) = {1 \over H^{-1} \sin^2 \theta + \cos^2 \theta} \qquad
    f(r) = 1 - {r_0^{7-p} \over r^{7-p}} \,.
 }
The remaining non-zero forms for the D2-F1 case are
\eqn{D2Fone}{
    A_1 &= e^{-\phi_0} \tan\theta \left(1 - \frac{D}{H}\right) dx_2  \cr
    A_3 &= e^{-\phi_0} \cos\theta \coth\alpha \left(1 - H^{-1}\right) dt \wedge dx_1 \wedge dx_2
\,,
}
and for the D5-F1 case
\eqn{D5Fone}{
    A_4 &= - e^{-\phi_0} \tan\theta \left(1 - \frac{D}{H}\right) dx_2 \wedge dx_3 \wedge dx_4 \wedge dx_5 \cr
    F_3 &= 2 e^{-\phi_0}\sp r_0^2 \cos \theta \cosh \alpha \sinh \alpha \sp d\Omega_3
\,.
}

Despite the differences in the solutions for D$p$-D$(p-2)$ and D$p$-F1, the thermodynamics in the parameter
space of ($r_0$, $\alpha$, $\theta$) are actually identical up to factors of the string coupling, which does not
affect the stability properties.  Hence we find precisely the same stability
conditions as in the D$p$-D($p-2$) case (\ref{FSC}).  However, for $\g
\neq 1$, the relationship between ($\alpha$, $\theta$) and the number
densities of D-branes differs from that of the F-strings.  We will comment on this further in section~\ref{SUGRA}.

\subsection{The D$p$-D($p-4$) Case}
\label{DPFOUR}

So far, all cases have involved non-extremal generalizations of BPS bound states whose mass follows a Pythagorean relationship: for example, $M = \sqrt{(N_2 V_2 \tau_2)^2 + (N_0 \tau_0)^2}$ for the D2-D0 bound state.  Intuitively, what tends to prevent a GL instability for this system is that the mass is a convex function of the D0-brane charge density.  This convexity becomes extremely weak in the limit where the D0-branes make the dominant contribution to the mass, so we are unsurprised to see the boundary of stability approach extremality in this limit.

It is also interesting to consider cases where the BPS bound state is at threshold.  For example, $M = N_4 V_4
\tau_4
+ N_0 \tau_0$ for the D4-D0 system, and a similar relation holds for D5-D1 and D6-D2.  The intuitive reasoning outlined above would lead us to expect that there is a GL instability for any amount of non-extremality.  We will see that this is the right answer for D6-D2 and D5-D1, but it is wrong for D4-D0.

The supergravity background solutions were given in \cite{peet} for the D5-D1 and D6-D2 systems.  Both of these cases,
as well as the D4-D0 solution, can be compactly written as
\eqn{DpDpminusFour}{
  ds_{\rm str}^2 &= \left(f_p f_{p-4}\right)^{-1/2} \left[-f dt^2 + f_{p-4} \left(dx_1^2 + \cdots + dx_4^2 \right) +
    \sum_{i=5}^p dx_i^2\right]
     \cr &\qquad{} +
    \left(f_p f_{p-4}\right)^{1/2} \left[f^{-1} dr^2 + r^2 d\Omega_{8-p}^2 \right]  \cr\noalign{\vskip2\jot}
  e^\phi &= e^{\phi_0} f_p^{(3-p)/4} f_{p-4}^{(7-p)/4}  \cr\noalign{\vskip2\jot}
    f(r) &= 1 - {r_0^{7-p} \over r^{7-p}} \qquad
    f_p(r) = 1 + {r_0^{7-p} \sinh^2 \alpha_p \over r^{7-p}} \qquad
    f_{p-4}(r) = 1 + {r_0^{7-p} \sinh^2 \alpha_{p-4} \over r^{7-p}}  \cr\noalign{\vskip2\jot}
   Q_p &= {(7-p) V_p \Omega_{8-p} \over 32 \pi G_N}\sp r_0^{7-p}  \sinh 2\alpha_p \qquad
  Q_{p-4} = {(7-p) V_p \Omega_{8-p} \over 32 \pi G_N}\sp r_0^{7-p}  \sinh 2\alpha_{p-4}
\,.
}
As before, there is a three-parameter family of solutions, which in this case is
parameterized by $(r_0,\alpha_p,\alpha_{p-4})$.  To compare with the D$p$-D($p-2)$ system, it
will be convenient to
define $\alpha$ and $\theta$ according to $\cos\theta \sinh 2\alpha = \sinh 2\alpha_p$ and
$\sin\theta \sinh 2\alpha = \sinh 2\alpha_{p-4}$: then $\tan\theta = Q_{p-4}/Q_p$, and $Q =
\sqrt{Q_{p-4}^2+Q_p^2} \propto r_0^{7-p} \sinh 2\alpha$, as in \thermo.  The expressions for the remaining thermodynamic variables are:
 \eqn{thermoDfour}{
  M &= {V_p \Omega_{8-p} \over 16 \pi G_N}
    {r_0^{7-p} \over 2} \left(2 + (7 - p) (\cosh 2\alpha_p + \cosh 2\alpha_{p-4})\right)  \cr
  T &= {7-p \over 4\pi r_0 \cosh\alpha_p \cosh\alpha_{p-4}} \qquad
    S = {V_p \Omega_{8-p} \over 4 G_N } r_0^{8-p} \cosh\alpha_p \cosh\alpha_{p-4}  \cr
  \mu_p &= \tanh\alpha_p \qquad \mu_{p-4} = \tanh\alpha_{p-4} \,.
 }
The resulting Hessian determinant is given by
\eqn{Hessfour}{
  \det H = { 8 G_N^2 \over \Omega_{8-p}^2 V_p^2 } {r_0^{2(p-8)}\sech^2 \alpha_p \sech^4 \alpha_{p-4}
\left(p-7 - (p-5)\cosh 2\alpha_p\right) \over
\cosh 2\alpha_{p-4} \left(7-p + 2\cosh 2\alpha_p \right) + (7-p) \cosh 2\alpha_p } \,.
}
Everything but the last factor in the numerator is positive definite.  For $p=6$, the
last factor is $-(1 + \cosh 2\alpha_p)$, and the system is
unstable all the way up to extremality, where $\det H = 0$.  For $p=5$, the last factor is simply $-2$, and we find a
result similar to $p=6$.  For $p=4$, however, the bound state is unstable when
 \eqn{DFourUnstable}{
  \cosh 2\alpha_4 < 3
 }
Consequently, if no D4 branes are present, then the D0 branes will clump as expected,
but if the D4-branes make a sufficiently dominant contribution to the total mass, the bound state is stable.

\begin{figure}
  \centerline{\includegraphics[width=3.5in]{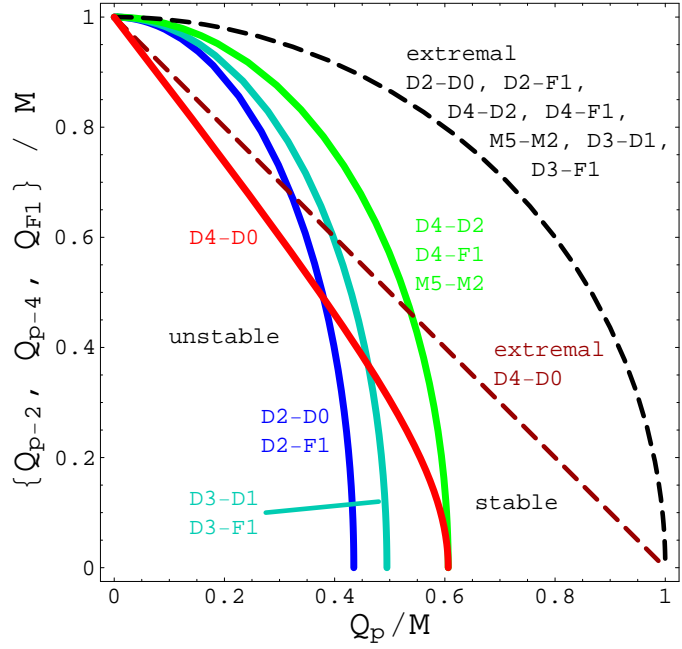}}
  \caption{The stability curves for various D-brane bound states are shown as solid curves.  The horizontal axis is the mass fraction coming from the branes of larger dimension.  The vertical axis is the mass fraction coming from the branes of smaller dimension.  In each case, the stable region is to the right of the solid curve.  The dashed curves correspond to extremal brane configurations.  The NCOS and OM theory limits are near the upper left corner.}\label{figA}
 \end{figure}

One may straightforwardly convert the inequalities \FSC\ and \DFourUnstable\ to relations between $Q_p/M$ and one of
the quantities $Q_{p-2}/M$, $Q_{p-4}/M$, and $Q_{F1}/M$, as appropriate.  The resulting stability curves are shown in
figure~\ref{figA}.  The stability curve for the D4-D0 bound state, for example, comes from setting $\cosh 2\alpha_4 = 3$, so the curve may be expressed parametrically in terms of $\alpha_0$:
 \eqn{DfourDzeroFinal}{
  {Q_4 \over M} = {3 \sqrt{8} \over 11 + 3 \cosh 2\alpha_0} \qquad
  {Q_0 \over M} = {3 \sinh 2\alpha_0 \over 11 + 3 \cosh 2\alpha_0}
 }

\section{Application to NCOS Theory and OM Theory}
\label{OMNCOS}

As an example of the utility of the CSC, we now consider its implications for theories of D-brane bound states.
Of particular interest is NCOS \cite{gukov,seiberg,gopakumar} theory, which consists fundamental strings on a stack of
D$p$ branes in a limit where the open strings decouple from gravity.  OM theory \cite{om,Bergshoeff:2000ai} is essentially the
11-dimensional lift of the D4-F1 NCOS system.

\subsection{NCOS Instabilities}
\label{SUGRA}

The NCOS limit is one in which the electric field living on a stack of D-branes approaches a critical value,
at which the effective string tension---defined as the energy per unit length needed to free a string from the bound
state---goes to zero relative to the actual string tension:
\eqn{TensionLimit}{
\frac{\alpha'}{\alpha'_{\rm eff}} = \frac{E_c^2 - E^2}{E_c^2} \rightarrow 0 \,.
}
The open string coupling is defined by a similar scaling:
\eqn{CouplingLimit}{
G_0^2 = \g \sqrt{\frac{E_c^2 - E^2}{E_c^2}} \,.
}
In the NCOS limit, $\g \rightarrow \infty$ with $G_0$ fixed.  For small $G_0$, NCOS exhibits
Hagedorn behavior with temperature \cite{GubserNCOS}:
\eqn{Htemp}{
T_H = \frac{1}{\sqrt{8\pi^2 \alpha'_{\rm eff}}} \,.
}
Finally, there is a relationship between the open string coupling, the number of F-strings per unit transverse
volume, and the number of D$p$-branes, $N_p$ \cite{GubserNCOS}:
\eqn{GONF}{
\frac{N_F}{V_t} = N_p \frac{1}{(2\pi \sqrt{\alpha'})^{p-1}} \frac{1}{G_0^2} \,.
}
We will choose units with $\alpha' = 1$.
Note that in the extremal limit, the BPS mass formula becomes
\eqn{BPS}{
M = \sqrt{(N_p V \tau_p)^2 + (N_F L \tau_f)^2} = \sqrt{N_p^2 \g^{-2} + N_F^2} \,,
}
where in the last equality we have chosen $V = (2\pi)^p$ and $L = 2\pi$.  The open string coupling also simplifies
with this choice, giving $G_0^2 = N_p / N_F$.

The supergravity background for NCOS is simply the solution described in the D$p$-F1 section above.  However, the
parameters ($r_0$, $\alpha$, and $\theta$) implicitly depend on $\g$.  In particular, in order
for the supergravity mass \thermo\ to match the BPS expression \BPS, we require\footnote{If instead
this were the D$p$-D($p-2$) bound state, the correct expression would be $\frac{(7-p)
\Omega_{8-p}}{(2\pi)^{7-p}} r_0^{7-p} \cosh\alpha
\sinh\alpha = \g \sqrt{N_p^2 + N_{p-2}^2 }$ since there is no relative factor of the string coupling in the
respective tensions.}:
\eqn{COSH}{
\frac{(7-p) \Omega_{8-p}}{(2\pi)^{7-p}} r_0^{7-p} \cosh\alpha \sinh\alpha = \g \sqrt{N_p^2 + N_F^2\sp \g^2} \,.
}
where we have chosen our conventions such that $ds_E^2 = e^{(\phi_0 - \phi)/2}
ds_{\rm str}^2$ and $16\pi G_N = (2\pi)^7 \alpha'^4 \g^2$.  In terms of the number of D-branes and strings,
$\theta$ can be written as:
\eqn{THETA}{
\sin\theta = \frac{N_F \g}{\sqrt{N_p^2 + N_F^2\sp \g^2}} \,.
}
This relation implies
\eqn{TANTHETA}{
\tan\theta = \g \frac{N_F}{N_p} = \sqrt{\frac{E_c^2}{E_c^2 - E^2}} \equiv \eta \,,
}
where we have used \CouplingLimit\ and \GONF.

From the stability curve plots, we see that we move toward the stable region as we approach extremality, \ie,
$T=0$.  For fixed $N_p$ and $N_F$, we can therefore find a critical temperature above which the system is
unstable.  We will replace $r_0$ in favor of $N_p$.  As usual, $\theta$ determines the ratio of the D$p$ and F1
contributions, and we will let $\alpha$ control the temperature.  We therefore have a critical value of $\alpha$---defined by $\sqrt{5-p} \sinh\alpha_c \cos\theta = 1$---below which the system is unstable.  Since we know
that $\tan\theta = \eta$, we can write the critical temperature in terms of $\alpha_c$ as
\eqn{AlphaBeta}{
\cosh\alpha_c = \sqrt{1 + \frac{1 + \eta^2}{5-p}} \approx \frac{\eta}{\sqrt{5-p}} \,,
}
where the approximate equality becomes exact in the limit of large $\eta$.

We can now put all the pieces together to find the critical temperature in the NCOS limit in terms of $N_p$,
the open string coupling $G_0$, and the scaling parameter $\eta$:
\eqn{CritTemp}{
T_c &= \frac{7-p}{4\pi r_0 \cosh\alpha_c} = \frac{7-p}{4\pi}\frac{1}{\cosh\alpha_c}
\left[\frac{(7-p) \Omega_{8-p} \cos\theta \cosh\alpha_c \sinh\alpha_c}{\g N_p (2\pi)^{7-p}}\right]^{1/(7-p)} \cr
&= \frac{7-p}{4\pi}\frac{1}{\cosh\alpha_c} \left[\frac{(7-p)
\Omega_{8-p}\cosh\alpha_c}{\g N_p (2\pi)^{7-p} \sqrt{5-p}}\right]^{1/(7-p)} \cr
&= \frac{7-p}{4\pi} \left[\frac{(7-p)\Omega_{8-p}}{N_p (2\pi)^{7-p} \sqrt{5-p}}\right]^{1/(7-p)}
\frac{1}{\cosh\alpha_c} \left[\frac{\cosh\alpha_c}{\g}\right]^{1/(7-p)} \cr
&= \frac{7-p}{4\pi} \left[\frac{(7-p)\Omega_{8-p}}{N_p (2\pi)^{7-p} \sqrt{5-p}}\right]^{1/(7-p)}
\sqrt{5-p} \left(\frac{1}{G_0^2 \sqrt{5-p}}\right)^{1/(7-p)} \eta^{-1} \,.
}
This quantity should be compared to the Hagedorn temperature, which is $T_H = 1/\sqrt{8 \pi^2 \eta^2}$:
\eqn{TempCompare}{
T_c / T_H = \frac{7-p}{\sqrt{8\pi^2}} \left[\frac{(7-p)\Omega_{8-p}}{
(5-p) N_p}\right]^{1/(7-p)}\sqrt{5-p} \left(\frac{1}{G_0^2}\right)^{1/(7-p)} \,.
}
The main qualitative point of interest is that the $\eta$ dependence cancels out entirely.  In the NCOS limit,
depending on the choice of parameters, one can have a GL transition above or below the Hagedorn transition.  But for
$p=5$ and any amount of non-extremality, there is always a GL instability, as long as we trust supergravity and the
CSC.\footnote{As in \cite{ssg-recent}, there is a possibility that the wavelength of the GL instability approaches
infinity in the limit that defines NCOS theory (that is, $\eta \to \infty$).}  For $p < 5$, the numerical factor in
\TempCompare\ is ${\cal O}(1)$, so
\eqn{TempCompareSimp}{
T_c / T_H \simeq (N_p G_0^2)^{1/(p-7)} \,.
}
Hence, if the open string coupling is small (which is required for a Hagedorn-like analysis anyway), $T_c \gg T_H$ in
general, unless one is considering a very large number of D$p$-branes.

By way of comparison, consider the D3-F1 bound state in the limit $N_3 \to \infty$, $\g \to 0$ with $\lambda \equiv \g N_3$ fixed, and $N_F$ fixed as well.  This is wholly different from the 3+1-dimensional NCOS limit because now the D3-branes dominate the mass of the extremal bound state.  Referring to figure~\ref{figA}, we see that a GL instability occurs when $M \gsim 2 Q_3$.  The temperature at the point of marginal stability can be computed as
\eqn{MaxTemp}{
T_c = \frac{1}{3^{3/8} \pi (2\pi \lambda)^{1/4}} + {
\cal O}\left(\frac{1}{N_3^4}\right) \simeq \frac{1}{7.5 \lambda^{1/4}}
}
in units where $\alpha'=1$.  When there are no F1's, this is the maximum possible temperature of a uniform D3-brane.  It's interesting that this can be greater or less than the Hagedorn temperature of type IIB strings, according to the value of $\lambda$.

\subsection{OM Theory Instabilities}

OM theory \cite{om} consists of a stack of $N$ M5-branes housing a large number of M2-branes.  The thermodynamics of this
system is described using \thermo\ by taking $p=4$.  If we choose a background 3-form field
strength of the form
\eqn{Hcrit}{
 H_{012} = M_p^3 \tanh \beta \,,
}
with $\beta$ some parameter, then the fluctuating open membranes decouple from gravity in the
``OM theory limit" $\beta \to \infty$, since the proper energy scale for the fluctuating M2's
goes like \eqn{Meff}{
 M_{\rm eff}^3 = 2 M_p^3 e^{-2\beta} \,.
}
We will choose $M_p = 1$.  For some choices of $r_0$, $\alpha$, and $\theta$, \FSC\ gives some critical temperature, above
which the system of D-branes becomes unstable.  We therefore need to express the parameters of OM theory in terms of our
standard supergravity parameters.

The M5 action contains a term \cite{m5action}:
\eqn{MFiveaction}{
 S \supset \mu_5 \int C_3 \wedge (H_3 + C_3) \,.
}
The analogous term of the M2 action is:
\eqn{MTwoaction}{
 S \supset \mu_2 \int C_3 \,.
}
Now assume that the M2 branes are extended in the 012 directions, and the M5 branes are extended in the 012345 directions.
Then $C_3$ can be gauged away in the 345 directions, and hence the second term in \MFiveaction\ vanishes.  Comparing
\MFiveaction\ and \MTwoaction\, we find $\mu_2 = H_{345} \mu_5$, and using the result from \cite{om}, we arrive at:
\eqn{mubeta}{
 \mu_2 = \mu_5 \sinh\beta \qquad \Longrightarrow \qquad \sinh\beta = \tan\theta \,.
}
As a consistency check, we note that the respective parameter ranges are $\beta \in [0,\infty)$ and $\theta \in [0,\pi/2]$.
Furthermore, this expression fits with our expectation that the OM theory limit should consist of a large number of M2-branes.

The condition for thermodynamic criticality for the M5-M2 case is $\cos\theta \sinh\alpha_c = 1$.  Furthermore,
\cite{harmark} tells us that
\eqn{N}{
 N \equiv \frac{r_0^3}{\pi} \cosh\alpha\sinh\alpha\cos\theta = \frac{r_0^3 \cosh\alpha_c}{\pi} \,,
}
where the first relation holds generally, and the second holds only at criticality.  The critical temperature is therefore
given by:
\eqn{Tc}{
 T_c^3 = \left({3 \over 4\pi r_0 \cosh\alpha_c} \right)^3 = \left({3 \over 4\pi}\right)^3 {1 \over \pi N}
 \left(\cosh\alpha_c \right)^{-2} = \left({3 \over 4\pi}\right)^3 {1 \over \pi N} {1 \over 1 + \cosh^2\beta} \,.
}
We now want to compare this expression with the characteristic energy scale of OM
theory \Meff\ in the large $\beta$ limit:
\eqn{limit}{
 \lim_{\beta \to \infty} \frac{T_c}{M_{\rm eff}} = \frac{3}{4\pi} \left(\frac{2}{\pi N}\right)^{1/3} \simeq \frac{1}{5 N^{1/3}}
}
As expected from the duality to NCOS theory, we find a non-zero
critical temperature in OM theory.  In contrast to
the general NCOS result, this critical temperature is always less than the characteristic energy scale $M_{\rm eff}$,
and in fact can be much less than this scale for a large number of coincident M5 branes.
We could have instead arrived an identical result by taking $p=4$ in \CritTemp\ along
with the following relation between the OM and NCOS parameters \cite{om}:
\eqn{OMNCOSTwo}{
M_{\rm eff} \sqrt{\alpha'_{\rm eff}} 2^{1/3} = \frac{1}{G_0^{2/3}} \,.
}

\section{Conclusions}
\label{CONCLUDE}

Our main results, namely the stability curves predicted by the CSC for the D2-F1, D3-F1, D4-F1, and D4-D0 bound states, are shown in figure~\ref{figA}.  Stability curves for some other bound states, such as D2-D0, D3-D1, D4-D2, and M5-M2, are related to one of the above cases.  It is notable that configurations related by T-duality have different stability curves, roughly because the process of compactifying and smearing changes the nature of the instabilities.

Given the significant loopholes in our extension of the arguments of \cite{reall} to the charged case, one may question the application of the CSC to the various
bound states that we have analyzed.  Briefly, our stance is that the detailed numerical
checks for the D2-D0 case in \cite{ssg-recent} make it seem very likely that the loopholes can be
closed, or that string theory for some reason does not take advantage of them.  We nevertheless
suspect that counter-examples may exist to the CSC for certain specially arranged interactions.
We hope to report further on these issues in the future.

The upper left corner of figure~\ref{figA} corresponds to NCFT, NCOS, or OM theory limits, which are interesting because they are believed to represent non-gravitational limits of string / M-theory which are not described by ordinary quantum field theories.  As is apparent from the figure, the GL instability tends to persist in these limits, at least as predicted by the thermodynamic properties of the bound states.  In NCOS theory, the critical temperature for a GL instability is a finite multiple of the Hagedorn transition temperature, indicating that there is an interesting competition between the tendency of strings to bunch up transversely on the branes (the GL effect) versus boiling off the branes (the Hagedorn effect).  As in the NCFT case, however, it is possible that the wavelength of the GL effect becomes large as one approaches the decoupling limit.  For OM theory the critical temperature is suppressed relative to the characteristic scale of open membranes by a power of the number of M5-branes, and similar speculations about wavelengths might be made.  The only way to check them at present is a rather difficult sort of numerics---difficult because the modes in question have slow decay at infinity, making it hard to isolate them in a shooting algorithm from uniform perturbations of the branes.

\section*{Acknowledgements}

We thank I.~Mitra for useful discussions.  J.F.~also thanks the
University of Wisconsin-Madison physics department for their
hospitality while this work was in progress.  This work was supported in
part by the Department of Energy under Grant No.\ DE-FG02-91ER40671, and
by the Sloan Foundation.  The work of J.F.~was also supported in part by
the NSF Graduate Research Fellowship Program.

\section*{Appendix}

For pure gravity, the existence of configurations with the properties described in and around \SpecialAction\ was demonstrated in \cite{reall, Prestidge, wy}.  In this case, for static configurations with the same symmetries as the uniform, on-shell, uncharged black brane, the equality \SpecialAction\ can be demonstrated using only the Hamiltonian constraint $G_{tt}=0$ together with boundary conditions at the horizon and at asymptotic infinity.  To be more precise: the action is
 \eqn{Ialways}{
  I_G = {1 \over 16\pi G} \int_M d^D x \, \sqrt{g} \, R + 
   {1 \over 8\pi G} \int_{\partial M} d^{D-1} x \, \sqrt{\gamma} \,
   (\Theta - \Theta_0) \,,
 }
where $\sqrt{\gamma} d^{D-1} x$ is the induced volume form on $\partial M$ and $\Theta = g^{\mu\nu} \Theta_{\mu\nu}$ is the trace of the extrinsic curvature tensor,
 \eqn{ExtrinsicCurvature}{
  \Theta_{\mu\nu} = (\delta^\lambda_\mu - n_\mu n^\lambda)
   \nabla_\lambda n_\nu \,,
 }
where $n_\mu$ is the outward pointing unit normal on $\partial M$.  The static ansatz is
 \eqn{BlackString}{
  ds^2 = -e^{2A} dt^2 + e^{2B} dx^2 + e^{2C} dr^2 + 
    e^{2D} d\Omega_{D-3}^2
 }
where the functions $A$, $B$, $C$, and $D$ depend only on $r$.\footnote{Actually, it is sometimes possible to prove \SpecialAction\ even when $A$ has both $r$ and $t$ dependence \cite{wy}.}  Assume that \BlackString\ satisfies $G_{tt}=0$.  Then the action \Ialways\ can be expressed purely in terms of an integral over the boundary $\partial M$, and, after rotation to Euclidean signature with periodic time, this integral expression reduces precisely to $I = \beta E - S$.  In the nomenclature of \cite{wy}, this is the reduced action.

It should be possible to incorporate locally conserved charges: some considerations in this direction can be found in \cite{hawkinghorowitz,hawkingross}.  The reasoning is as follows.  Let a black string carry electric charge under a gauge field $C_{(1)} = C_\mu dx^\mu$ (higher-dimensional cases follow a similar pattern of reasnoning), whose action is 
 \eqn{Iform}{
  I_{C} = -{1 \over 4} \int d^D x \, \sqrt{g} \, F_{\mu\nu}^2 \,.
 }
The metric ansatz is as in \BlackString, and in addition one assumes that the only non-zero components of the gauge field are $C_t(r)$ and $C_r(r)$.  The Hamiltonian constraints are now $G_{tt} = 8\pi G T_{tt}$ and $\nabla_i E^i = 0$, where $E^i$ is the momentum conjugate to $C_i$, so that $E^i = F^{0i}$ on-shell---but this last equation is {\it not} part of the Hamiltonian constraints.  Evaluating $I = I_G + I_C$ on the ansatz described, subject to the Hamiltonian constraints, and performing a rotation to Euclidean signature with periodic time, leads again to \SpecialAction, where now $B_1 = \mu = C_t(r_H) - C_t(\infty)$ is the gauge potential at the horizon and $A_1$ is the electric charge
 \eqn{ExpressQ}{
  Q = \int_{S^{D-2}} d^{D-2} x \, \sqrt{h} \, n_i E^i \,,
 }
where $\sqrt{h} d^{D-2}x$ is the volume form induced on $S^{D-2}$ and $n_i$ is the outward-pointing unit normal (normal also to the time direction).  By virtue of the Gauss law constraint, the $S^{D-2}$ can be any sphere that entirely encloses the horizon.

Evidently, in these examples, there is more than a finite-dimensional space of deformations of the on-shell solution satisfying \SpecialAction: $A(r)$ remains an arbitrary function, as does $C_t(r)$.  We do not have a general proof that this will always be so, nor have we shown that $\partial A_j/\partial a_i$ is non-singular for some appropriately chosen $n$-parameter family of deformations---though this last property can usually be achieved by having $a_i = A_i$.  We hope however that the considerations presented here make more plausible the claims made around \SpecialAction.  The bottom line is that this integrated form of the first law only employs symmetries, certain boundary conditions, and the small subset of the equations of motion comprising the constraints in a Hamiltonian framework---so it is sensible to think that there is a fairly broad class of ``partially on-shell'' configurations (i.e.~configurations satisfying the constraints) to which it applies.

\bibliographystyle{ssg}
\bibliography{paper}

\begingroup\raggedright\begin{thebibliography}{10}

\bibitem{glOne}
R.~Gregory and R.~Laflamme, ``Black strings and p-branes are unstable,'' {\em
  Phys. Rev. Lett.} {\bf 70} (1993) 2837--2840,
  \href{http://xxx.lanl.gov/abs/hep-th/9301052}{{\tt hep-th/9301052}}.

\bibitem{glTwo}
R.~Gregory and R.~Laflamme, ``The Instability of charged black strings and
  p-branes,'' {\em Nucl. Phys.} {\bf B428} (1994) 399--434,
  \href{http://xxx.lanl.gov/abs/hep-th/9404071}{{\tt hep-th/9404071}}.

\bibitem{gmOne}
S.~S. Gubser and I.~Mitra, ``Instability of charged black holes in anti-de
  Sitter space,'' \href{http://xxx.lanl.gov/abs/hep-th/0009126}{{\tt
  hep-th/0009126}}.

\bibitem{gmTwo}
S.~S. Gubser and I.~Mitra, ``The evolution of unstable black holes in anti-de
  Sitter space,'' {\em JHEP} {\bf 08} (2001) 018,
  \href{http://xxx.lanl.gov/abs/hep-th/0011127}{{\tt hep-th/0011127}}.

\bibitem{reall}
H.~S. Reall, ``Classical and thermodynamic stability of black branes,'' {\em
  Phys. Rev.} {\bf D64} (2001) 044005,
  \href{http://xxx.lanl.gov/abs/hep-th/0104071}{{\tt hep-th/0104071}}.

\bibitem{ssg-recent}
S.~S. Gubser, ``The Gregory-Laflamme instability for the D2-D0 bound state,''
  \href{http://xxx.lanl.gov/abs/hep-th/0411257}{{\tt hep-th/0411257}}.

\bibitem{gukov}
S.~Gukov, I.~R. Klebanov, and A.~M. Polyakov, ``Dynamics of (n,1) strings,''
  {\em Phys. Lett.} {\bf B423} (1998) 64--70,
  \href{http://xxx.lanl.gov/abs/hep-th/9711112}{{\tt hep-th/9711112}}.

\bibitem{seiberg}
N.~Seiberg, L.~Susskind, and N.~Toumbas, ``Strings in background electric
  field, space/time noncommutativity and a new noncritical string theory,''
  {\em JHEP} {\bf 06} (2000) 021,
  \href{http://xxx.lanl.gov/abs/hep-th/0005040}{{\tt hep-th/0005040}}.

\bibitem{gopakumar}
R.~Gopakumar, J.~M. Maldacena, S.~Minwalla, and A.~Strominger, ``S-duality and
  noncommutative gauge theory,'' {\em JHEP} {\bf 06} (2000) 036,
  \href{http://xxx.lanl.gov/abs/hep-th/0005048}{{\tt hep-th/0005048}}.

\bibitem{om}
R.~Gopakumar, S.~Minwalla, N.~Seiberg, and A.~Strominger, ``OM theory in
  diverse dimensions,'' {\em JHEP} {\bf 08} (2000) 008,
  \href{http://xxx.lanl.gov/abs/hep-th/0006062}{{\tt hep-th/0006062}}.

\bibitem{Ross:2005vh}
S.~F. Ross and T.~Wiseman, ``Smeared D0 charge and the Gubser-Mitra
  conjecture,'' \href{http://xxx.lanl.gov/abs/hep-th/0503152}{{\tt
  hep-th/0503152}}.

\bibitem{Bostock:2004mg}
P.~Bostock and S.~F. Ross, ``Smeared branes and the Gubser-Mitra conjecture,''
  {\em Phys. Rev.} {\bf D70} (2004) 064014,
  \href{http://xxx.lanl.gov/abs/hep-th/0405026}{{\tt hep-th/0405026}}.

\bibitem{AharonyEtAlOne}
O.~Aharony, J.~Marsano, S.~Minwalla, and T.~Wiseman, ``Black hole - black
  string phase transitions in thermal 1+1 dimensional supersymmetric Yang-Mills
  theory on a circle,'' \href{http://xxx.lanl.gov/abs/hep-th/0406210}{{\tt
  hep-th/0406210}}.

\bibitem{Harmark:2004ws}
T.~Harmark and N.~A. Obers, ``New phases of near-extremal branes on a circle,''
  {\em JHEP} {\bf 09} (2004) 022,
  \href{http://xxx.lanl.gov/abs/hep-th/0407094}{{\tt hep-th/0407094}}.

\bibitem{harmark}
T.~Harmark and N.~A. Obers, ``Phase structure of non-commutative field theories
  and spinning brane bound states,'' {\em JHEP} {\bf 03} (2000) 024,
  \href{http://xxx.lanl.gov/abs/hep-th/9911169}{{\tt hep-th/9911169}}.

\bibitem{Harmark:2000ff}
T.~Harmark, ``Open branes in space-time non-commutative little string theory,''
  {\em Nucl. Phys.} {\bf B593} (2001) 76--98,
  \href{http://xxx.lanl.gov/abs/hep-th/0007147}{{\tt hep-th/0007147}}.

\bibitem{roy}
J.~X. Lu and S.~Roy, ``Non-threshold (f,D p) bound states,'' {\em Nucl. Phys.}
  {\bf B560} (1999) 181--206,
  \href{http://xxx.lanl.gov/abs/hep-th/9904129}{{\tt hep-th/9904129}}.

\bibitem{Harmark:2000wv}
T.~Harmark, ``Supergravity and space-time non-commutative open string theory,''
  {\em JHEP} {\bf 07} (2000) 043,
  \href{http://xxx.lanl.gov/abs/hep-th/0006023}{{\tt hep-th/0006023}}.

\bibitem{peet}
A.~W. Peet, ``TASI lectures on black holes in string theory,''
  \href{http://xxx.lanl.gov/abs/hep-th/0008241}{{\tt hep-th/0008241}}.

\bibitem{Bergshoeff:2000ai}
E.~Bergshoeff, D.~S. Berman, J.~P. van~der Schaar, and P.~Sundell, ``Critical
  fields on the M5-brane and noncommutative open strings,'' {\em Phys. Lett.}
  {\bf B492} (2000) 193--200,
  \href{http://xxx.lanl.gov/abs/hep-th/0006112}{{\tt hep-th/0006112}}.

\bibitem{GubserNCOS}
S.~S. Gubser, S.~Gukov, I.~R. Klebanov, M.~Rangamani, and E.~Witten, ``The
  Hagedorn transition in non-commutative open string theory,'' {\em J. Math.
  Phys.} {\bf 42} (2001) 2749--2764,
  \href{http://xxx.lanl.gov/abs/hep-th/0009140}{{\tt hep-th/0009140}}.

\bibitem{m5action}
I.~A. Bandos {\em et.~al.}, ``Covariant action for the super-five-brane of
  M-theory,'' {\em Phys. Rev. Lett.} {\bf 78} (1997) 4332--4334,
  \href{http://xxx.lanl.gov/abs/hep-th/9701149}{{\tt hep-th/9701149}}.

\bibitem{Prestidge}
T.~Prestidge, ``Dynamic and thermodynamic stability and negative modes in
  Schwarzschild-anti-de Sitter,'' {\em Phys. Rev.} {\bf D61} (2000) 084002,
  \href{http://xxx.lanl.gov/abs/hep-th/9907163}{{\tt hep-th/9907163}}.

\bibitem{wy}
B.~F. Whiting and J.~W. York, ``Action principle and partition function for the
  gravitational field in black hole topologies,'' {\em Phys. Rev. Lett.} {\bf
  61} (1988) 1336.

\bibitem{hawkinghorowitz}
S.~W. Hawking and G.~T. Horowitz, ``The Gravitational Hamiltonian, action,
  entropy and surface terms,'' {\em Class. Quant. Grav.} {\bf 13} (1996)
  1487--1498, \href{http://xxx.lanl.gov/abs/gr-qc/9501014}{{\tt
  gr-qc/9501014}}.

\bibitem{hawkingross}
S.~W. Hawking and S.~F. Ross, ``Duality between electric and magnetic black
  holes,'' {\em Phys. Rev.} {\bf D52} (1995) 5865--5876,
  \href{http://xxx.lanl.gov/abs/hep-th/9504019}{{\tt hep-th/9504019}}.

\end{thebibliography}\endgroup
\end{document}